\documentclass[prb,twocolumn,showpacs]{revtex4}% Physical Review B
\usepackage{graphicx}% Include figure files
\usepackage{dcolumn}% Align table columns on decimal point
\usepackage{natbib}
\usepackage[english]{babel}
\usepackage{epsfig}

\begin{document}
\newcommand{\putgraph}[2]{\includegraphics[#1]{#2.jpg}} % for pdflatex

\newcommand{\fig}[1]{Fig.(\ref{fig:#1})}
\newcommand{\figa}[2]{Fig.(\ref{fig:#1}#2)}
\newcommand{\figas}[3]{Fig.(\ref{fig:#1}#2-#3)}
\newcommand{\figs}[2]{Figs.(\ref{fig:#1}-\ref{fig:#2})}
\newcommand{\tab}[1]{Tab.(\ref{tab:#1})}
\newcommand{\tabs}[2]{Tabs.(\ref{tab:#1}-\ref{tab:#2})}
\newcommand{\tabb}[2]{Tabs.(\ref{tab:#1}) and (\ref{tab:#2})}
\newcommand{\bom}{$\mu_{B}$~}
\newcommand{\eq}[1]{Eq.(\ref{#1})}
\newcommand{\eqs}[2]{Eq.(\ref{#1}-\ref{#2})}
\newcommand{\sct}[1]{Sec.(\ref{sec:#1})}

\newcommand{\sigmab}{$\sigma_{pz}$~}
\newcommand{\sigmaab}{$\sigma_{pz}^{*}$~}
\newcommand{\pib}{$\pi_{p}$~}
\newcommand{\piab}{$\pi_{p}^{*}$~}
\newcommand{\nn}{N$_2$}
\newcommand{\bb}{B$_2$}

\newcommand{\ang}{\AA$^{-2}$~}
\newcommand{\mub}{$\mu_B$~}
\newcommand{\abit}{\emph{ab initio}~}
\newcommand{\abitt}{\emph{ab initio}~}
\newcommand{\abz}{A$_{BZ}$}
\newcommand{\etal}{\emph{et al.}}

\newcommand{\cel}{C$^\circ$}

\newcommand{\rea}{X$^*$+Y$^*$$\rightarrow$XY$^*$}

\newcommand{\sta}{$^*$}

\newcommand{\bbrea}{B$^*$+B$^*\rightarrow$B$_2^*$}
\newcommand{\bnrea}{B$^*$+N$^*\rightarrow$BN$^*$}

\newcommand{\refcite}[2]{#1.[\onlinecite{#2}]} % APS
% \newcommand{\refcite}[2]{#1.\cite{#2}} % elsevier

% \begin{frontmatter} % elsevier

\title{Computational study of boron nitride nanotube synthesis: how catalyst
morphology stabilizes the boron nitride bond.}

% ----------- APS style authors ---------------------------
\author{S. Riikonen,$^1$ A.~S. Foster,$^{1,2}$ A.~V. Krasheninnikov,$^{1,3}$
and R.~M. Nieminen$^{1}$}
\email{sampsa.riikonen@iki.fi}

\affiliation{$^1$COMP/Department of Applied Physics, Helsinki University of
Technology, P.O. Box 1100, FI-02015, Finland}

\affiliation{$^2$Department of Physics, Tampere University of Technology
P.O. Box 692, FI-33101 TUT, Tampere, Finland}

\affiliation{$^3$ Materials Physics Division, University of Helsinki,
P.O. Box 43, FI-00014,  Finland}
% --------------------------------------------------------------------
%
% ---------- elsevier type authors -------------------
% \author[fyslab]{S. Riikonen}
% \author[fyslab,tampere]{A.~S. Foster}
% \author[fyslab,helsinki]{A.~V. Krasheninnikov}
% \author[fyslab]{R.~M. Nieminen}
% \address[fyslab]{COMP/Department of Applied Physics, Helsinki University of Technology, P.O. Box 1100, FI-02015, Finland}
% \address[tampere]{Department of Physics, Tampere University of Technology
% P.O. Box 692, FI-33101 TUT, Tampere, Finland}
% \address[helsinki]{Materials Physics Division, University of Helsinki,
% P.O. Box 43, FI-00014,  Finland}

\begin{abstract}
In an attempt to understand why
catalytic methods for the growth of
boron nitride nanotubes work much worse than for their carbon counterparts,
we use first-principles calculations to
study the energetics of elemental reactions forming \nn, \bb~and BN molecules on an
iron catalyst.  We observe that in the case of these small molecules, the catalytic activity is
hindered by the formation of \bb~ on the iron surface. We also observe that the local morphology of 
a step edge present in our nanoparticle model stabilizes the boron nitride molecule with
respect to \bb~due to the ability of the step edge to offer sites
with different coordination simultaneously for nitrogen and
boron. Our results emphasize the importance of atomic steps for
a high yield chemical vapor deposion growth of BN nanotubes and may
outline new directions for improving the efficiency of the method.
\end{abstract}

 % APS

\maketitle % aps

% \end{frontmatter} % elsevier

\section{Introduction}
\label{sec:intro}
Boron nitride nanotubes (BNNT) consist of hexagonal graphitic-like sheet of
alternating boron and nitrogen atoms rolled into a tube
\cite{chopra95,golberg96,rubio94}.
The structure of BNNTs is analogous to the more well-known
(monatomic) carbon nanotubes (CNT), but their physical properties
are quite different from those of their carbon counterpart. The
mechanical and wear-resistant properties of both materials are of
the same impressive order (for example, the Young's modulus is in
the terapascal range \cite{hernandez98}), while the electronic
properties of BNNTs can be more attractive.  CNTs are either metals
or semiconductors depending on their chirality, while BNNTs are
always semiconductors \cite{blase94,charlier97} with the gap ($\sim$
5.5 eV) practically independent of the nanotube chirality and its
diameter \cite{blase94}.  As hexagonal boron nitride (h-BN) is very
resistant to oxidation\cite{paine90,tang03}, BNNTs which
inherit these properties, are suitable for shielding and coating at
the nanoscale.  Despite these prospects, BNNTs have received very
little attention compared to CNTs due to various difficulties in
their reproducible and efficient synthesis \cite{golberg07}.

The fact that the BNNT consists of two different atomic species
implies that the synthesis of BNNTs is more complicated than the
synthesis of monatomic CNTs, as additional chemical reactions are
possible.  CNTs are typically synthesized from hydrocarbon
precursors \cite{grobert07,terranova06} and according to current
theoretical understanding of the CNT formation process, individual
carbon atoms diffuse in or on a metal nanoparticle, forming
graphitic networks that eventually gives rise to the appearance of a
CNT (see e.g. \refcite{Refs}{raty05,abildpedersen06,amara09}).

Assuming that these ideas are relevant to the growth of BNNTs, it
becomes important to understand the factors that determine whether
individual nitrogen and boron atoms diffusing on a catalytic surface
result in the formation of BN structures, or \nn~molecules and B
clusters.

In this paper, in an attempt to understand why catalytic methods for
the growth of BNNTs work much worse than for their carbon
counterparts, we use first-principles calculations to study the
behavior of \nn, \bb~and BN molecules on an iron catalyst. Such
molecules are the simplest systems involved, and the complete
understanding of their behavior on the catalyst surface is a
prerequisite to understanding the whole process. We assume an ideal
situation, where the precursors used for producing BNNTs (and
similar structures), are decomposed into individual boron and
nitrogen atoms and deposited on the catalyst. We chose iron as the
typical catalyst used in chemical vapor deposition (CVD) growth.
We then investigate under which situations the BN formation becomes
energetically favorable. We show that on a (110) close-packed
surface of BCC iron,  \bb~formation will dominate while at step edge
regions, BN formation will be the most favorable reaction.

This paper is organized as follows: In \sct{bntsyn} we first give a
brief review of the synthesis methods of BNNTs and similar
structures.  In \sct{setup} we explain the approximations and
the computational approach we have chosen and how they can be justified. 
In \sct{methods} we discuss in detail the 
computational methods.  In \sct{results}, we present our
results and demonstrate how specific catalyst morphologies stabilize
the BN bond. To better understand the underlying chemistry,
in \sct{ele} we analyze the electronic structure of the adsorbed molecules. 
Finally in (\sct{conclusions}), we discuss how BN catalytic synthesis on iron might be
spoiled and how the situation could be improved.

\section{Synthesis of BNNTs and related structures}
\label{sec:bntsyn}
BNNTs have been synthesized with various methods and in a
wide range of temperatures. Nearly all the methods show traces of 
metal particles, but their role as a catalyst is far from clear.
In this section, we give a brief overview of BNNT synthesis, with the emphasis on
the role of catalysts if present in the synthesis method.

\subsection{Arc-discharge}
BNNTs were synthesized for the first time with the arc-discharge method, using
BN-packed tungsten anode and copper cathode \cite{chopra95}.
Successively various anode and cathode materials, including hafnium diboride\cite{loiseau96}, 
tantalum press-filled with boron nitride \cite{terrones96} and a mixture of
boron, nickel and cobalt \cite{cumings00} have been used.
Typically, amorphous particles have been observed at the BNNT tips \cite{terrones96} or encapsulated in BN cages \cite{loiseau96}. These particles could be metallic (borides), implying a metal catalyzed synthesis \cite{terrones96}, while the encapsulated material could also be BN and the synthesis would be non-catalytic \cite{loiseau96}.
A non-catalytic open-ended growth (involving no nanoparticles) has also been proposed \cite{cumings00}.

Keeping in mind that temperatures in the arc-discharge method reach beyond 3000 \cel,
it is probably not well-suited for mass production of BNNTs.

\subsection{Laser-ablation}
The laser ablation method is based on the Vapour-Liquid-Solid (VLS) model \cite{morales98}, in which
the target material is evaporated and precipitated from the vapor-phase, eventually forming nanoparticles
and solid, wire-like nanostructures.  These are then carried by a gas flow to a collector \cite{morales98}.

Yu, Zhou, \etal \cite{zhou99,yu98} used BN powder as the target (T $\sim$ 1200 \cel) and observed that adding small amounts of catalyst Ni and Co into the target, resulted in longer nanotubes of better quality that were more often single-walled\cite{zhou99}. Metal particles were observed to encapsulate inside BN material and they were thought to play an important role in the synthesis \cite{yu98}.

In other studies featuring higher temperatures\cite{lee01,arenal07} (2400\cel-3000\cel), pure BN targets were used and BNNT growth from pure boron nanoparticles was observed\cite{lee01,arenal07}.

In other laser-based techniques used for synthesizing BNNTs, the resulting product is typically
collected directly from the target itself:  Laude et al. \cite{laude00} achieved BN dissociation by laser heating in low pressure nitrogen atmosphere.  This resulted in BNNTs and BN polyhedra that grew out of liquid boron \cite{laude00}.  Golberg et. al. \cite{golberg96} heated cubic BN by laser \cite{golberg96} in diamond anvil cell at high temperature and pressure, producing BNNTs directly from the liquid phase \cite{golberg96}.  Ablation of BN by high-frequency laser in low-pressure nitrogen atmosphere \cite{golberg03}, produced BNNTS and BN "nanohorns".

\subsection{Ball-milling and Annealing}
Annealing methods have been used to produce BN nanowires, "nanobamboos" and BNNTs.
These methods produce tubular BN structures by first milling the
boron containing starting material into a fine powder during long times (typically $\sim$ 24h)
and then annealing it at temperatures of $\sim$ 1000-1200\cel~in an inert \cite{bae03}
or nitrogen containing \cite{chen99,chen99_2,velazquezsalazar05,yu05,yu07} atmosphere. 
As the starting material, h-BN \cite{chen99,velazquezsalazar05,bae03} or pure boron
powder \cite{chen99_2,bae03,yu05,yu07} have been used.  During the milling,
the starting material can be activated \cite{yu07}, by performing the milling in
reactive atmospheres. %$\sim$ 300 kPa 
Pressurized N$_2$\cite{chen99} or
ammonia gases\cite{chen99_2,yu05,yu07} have been used for this purpose. 
%In other works \cite{bae03,velazquezsalazar05}, passive atmospheres have been used.
Nanosized metal particles observed frequently in the samples come from the metal balls used in the milling process.

There seems to be no generally accepted scheme how nanotubules form in this synthesis method.
%Metallic nanoparticles were observed frequently in the samples and sometimes it is concluded that they
%aid the growth of nanotubules \cite{bae03,velazquezsalazar05}, sometimes their role is not clear %\cite{chen99,chen99_2,yu07}
%and in some cases they play no role at all \cite{yu05}.
Metallic nanoparticles were observed frequently in the samples, and  it was argued 
that they facilitate the growth of nanotubules \cite{bae03,velazquezsalazar05}, while
it was concluded in other works that they are not important \cite{yu05}.  Some authors
simply state that their role is not clear\cite{chen99,chen99_2,yu07}.
% but their role is not clear
%\cite{bae03,velazquezsalazar05,chen99,chen99_2,yu05,yu07}.
In general, the nanotubes synthesized by these methods are of poor quality and the yields are very small,
so the methods are not, at least at the present stage, very suitable for mass-production of BNNTs.

Related to these methods is the work of Koi, Oku and co-workers \cite{tokoro04,koi05,oku07,narita06} in which either hematite\cite{tokoro04} or Fe$_4$N powder\cite{koi05,oku07,narita06}
together with boron powder was annealed in nitrogen atmosphere at $\sim$ 1000 \cel.  Iron particles coated in BN layers\cite{tokoro04,koi05}, BN nanowires\cite{koi05},hollow cages \cite{oku07}, "nanobamboo" structures \cite{narita06}, nanotubes and "cup-stacked" nanotubes \cite{oku07} were synthesized. In these works, the formation of BN layers in the reactions involving Fe$_4$N has been described in two
different ways:  Either Fe$_4$N and Fe$_2$B become liquid, boron segregates on the nanoparticle
surface and reacts with the N$_2$ atmosphere\cite{koi05}, or an amorphous boron layer on the Fe$_4$N is
converted to BN as the Fe$_4$N is reduced from nitrogen\cite{narita06}.

\subsection{Chemical Vapour Deposition}
In a Chemical Vapour Deposition (CVD) method, one or more volatile precursors react and decompose on 
the catalyst to form the desired compound.  CVD methods for producing BN filaments and BNNTs have been utilized in several works \cite{gleize94,shelimov00,lourie00,ma01_2,huo02,tang02_2,fu04}. 

Gleize \etal \cite{gleize94} used diborane and ammonia or N$_2$ gases as the boron and nitrogen containing precursors.  These were deposited on various boride surfaces (including Zr,Hf,Ti,V,Nb and Ta borides) at a temperature of 1100\cel.  It was observed that diborane did not play any role in the tubule growth (diborane and ammonia formed amorphous BN only), but the boron in the reaction came from the boride catalyst itself \cite{gleize94}.  The boride then acted both as a catalyst and as a reactant for the tubules.  Successive studies using similar temperatures have made the same observation.

Lourie \etal \cite{lourie00} deposited borazine on cobalt, nickel, and nickel boride catalyst particles
and concluded that the boride catalyst gave the best results.
Huo, Fu, \etal \cite{huo02,fu04} used for the nitrogen containing precursor a mixture of ammonia and nitrogen gas.  The boron source was again the catalyst itself which consisted of iron boride nanoparticles.
%Lowering the boron content in the nanoparticles down to $\sim$ 30 \%, resulted in enhanced BNNT vs. BN %nanowire production \cite{huo02,fu04}.

In another study \cite{tang02_2} nickel boride nanoparticles supported on alumina (in order to avoid nanoparticle agglomeration) with ammonia and nitrogen were used. BNNTs were observed to grow out of the nickel boride nanoparticles at T=1100-1300 \cel, while no "nanobamboo" structures were observed (anglomeration was avoided).

Ma \etal~emphasized that CVD using metal catalysts must be difficult due to the poor
wetting property of BN with metals \cite{ma01_2}.  For this reason they used melamine diborate to create
a metal-free B-N-O precursor \cite{ma01_2,ma01,ma02}.  This precursor then reacted with N$_2$ at 1200-1700 \cel.
Tip-growth of multi-walled BNNTs from amorphous B-N-O clusters was observed\cite{ma01_2}.  The synthesis was explained by condensation of BN from the vapor-phase into the B-N-O particles \cite{ma01_2}, or either
by reduction of B$_2$O$_3$ vapor \cite{ma01}.

Borazine and similar molecules have been used in CVD to produce BN nanotubules.
Shelimov and Moskovits \cite{shelimov00} created BN nanotubules by depositing 2,4,6-trichloroborazine on aluminum oxide at a temperature of 750 \cel.  These kinds of methods are based on the thermal decomposition (pyrolysis) of borazine and similar molecules on surfaces \cite{nagashima95} and there is a direct connection to the CVD synthesis of h-BN thin films, a theme that has been reviewed by Paine and Narula\cite{paine90}.

\subsection{Other}
Other methods include the substitution of carbon atoms in CNTs by boron and nitrogen \cite{han98, han99, golberg99, golberg00}, reduction-nitridation reactions \cite{chen04} and boric-acid reacting with activated carbon \cite{deepak02}.
Finally, the most succesful method up to date for synthesizing BNNTs is by Tang and co-workers \cite{tang02,zhi05,golberg07}.

In the method of Tang et. al. \cite{tang02,zhi05,golberg07}, boric oxide 
vapour was created in situ and reacted with ammonia
at temperatures T $\ge$ 1100 \cel.  Boric oxide was created from magnesium oxide and boron powder.
Magnesium was also thought to act as a catalyst in the reduction of boric oxide into boron nitride \cite{tang02}.
This method seems to be related to the ``classical high-temperature'' methods to produce bulk h-BN \cite{paine90},
where the formation of h-BN is attributed to the gas forming property of the undesired elements (oxygen) and the thermodynamical stability of h-BN \cite{paine90}.

By this method, boron and nitrogen could be converted into BNNTs by an efficiency of 40$\%$ \cite{tang02}
and hundreds of milligrams of BNNTs were produced.
Most of the nanotubes were open-ended, although some encapsulated material was found in the samples \cite{tang02}.
Liquid-phase magnesium drops could have catalyzed the reaction, but in this case they were evaporated in the final process \cite{tang02}.  The quantity and quality of BNNTs depended strongly on the temperature: below 1100 \cel, quality was better, but yield was small \cite{zhi05}.  Increasing the temperature, increased the yield, but tube diameter started to grow and BN flakes were formed when temperature was beyond 1250 \cel \cite{zhi05}.  Adding FeO to the initial MgO powder, solved this problem and BNNTs could be produced up to 1700 \cel \cite{zhi05}.  The growth then seemed to be catalytic \cite{zhi05}.

\subsection{Common features and the role of catalyst as the simulation challenge}
\label{sec:setup}
As evident from this brief review, BNNT nanotubes can be synthesized by various methods,
and in nearly all of them, metal particles which may have catalytic activity, are present.
However, the role of metal catalysts in BNNT growth is not well understood.
%In every class of synthesis, there are traces of metallic nanoparticles
%encapsulated in the samples. Sometimes they are concluded to contribute to
%the BNNT growth and sometimes not.
%The BNNT synthesis seems to have a
%very different nature in the different methods and in some cases the BN nanotubules
%are thought to grow out of liquid boron or recrystallization of amorphous boron is offered
%as the explanation, just to name few explanations.

In the CVD methods and when metal catalysts are involved, it seems to be important to use
borides instead of pure metals.  Borides are able to dissolve boron and nitrogen at the same time
\cite{gleize94}, while the solubility of boron for example in iron, is known to be very small
\cite{guo03}.  On the other hand, borides likely provide boron atoms during the BNNT
growth\cite{gleize94}, so they act both as the catalyst and the reactant itself, which is conceptually
very different from the case of CNT synthesis.

In methods using borazine and similar molecules, we must keep in mind that
these molecules already contain the desired boron nitride bonds.  We can then imagine
that the pyrolysis of these molecules in temperatures of T$\sim$800\cel~is used rather to
remove the hydrogen atoms, than breaking the boron nitride bonds.  This synthesis can then be
conceptually quite different from the other synthesis methods. 
Finally, in the state of the art method (Tang, Golberg, Zhi and others), the catalytic role of iron and magnesium used in the process is not fully understood.

All these synthesis methods pose interesting challenges for
theoretical calculations.  However, to our knowledge only a single \abit study
on BNNT synthesis has been published \cite{charlier99}.  In that study, the non-catalytic
growth of BNNTs was considered and it was shown that open-ended growth of single-walled 
armchair BNNTs is in principle possible \cite{charlier99}.

%Trying to solve any of the open questions posed by the experimental methods reviewed here,
%can be quite a complicated issue.

Modelling a catalytic process is a very challenging problem.
Many of the \abit studies in this field concentrate
in studying situations where the catalyst is reactive enough to dissociate
a precursor, while not being too reactive
to block the synthesis\cite{norskov02}.  A typical example of a thoroughly studied catalytic synthesis
process is the ammonia synthesis and its rate limiting step, the \nn~dissociation \cite{honkala05}.

In this work, we study the adsorption energies, reaction energies and some reaction barriers for simple boron and nitrogen containing molecules on a catalyst.  We are trying to find reasons why BNNT synthesis on transition metals has proven to be so difficult and if the boron nitride formation could be made energetically
favorable.  We do this by studying the stability of the boron nitride bond on iron.
This can be seen as a natural first step before addressing more complicated issues and catalysts (such as borides).

Our computational set up mimicks the CVD synthesis.  
We assume that the precursors (not defining them) have dissociated and donated 
B and N atoms on the catalyst.  In the simulations, we then adsorb individual B and N atoms
on the surface and calculate the reaction energetics when these adsorbed atoms (X$^*$ and Y$^*$)
form adsorbed molecular species (XY$^*$).

Thinking in terms of this simplified model of CVD synthesis it is easy to understand why 
boron nitride structures can be much more difficult to form than pure carbon structures;
in the carbon case and looking at the most simple molecules, we have only carbon atoms involved in the reactions i.e. XY\sta = C$_2$\sta, while in the boron nitride case we have several competing diatomic molecules, i.e. XY\sta = \nn\sta,~\bb\sta,~or BN\sta.

As the adsorbed boron and nitrogen atoms react on the catalyst surface, complicated 
surface species might form, for example, boron clusters, 
boron-iron clusters, BN molecules and chains and clusters consisting of both boron and nitrogen, etc.
If our goal is to understand the problems in BNNT synthesis in such a complex situation, a good first step is to study the most simple surface species, i.e. the adsorbed diatomic molecules that can be formed with adsorbed B and N.  If, by studying these simple diatomic molecules, we find situations where the catalyst ``promotes''   the formation of BN\sta~molecule instead of \nn\sta~and \bb\sta~molecules, this should have
consequences in more realistic situations as well.

Finally, we emphasize that in this work we are interested in the theoretical aspect
of the boron nitride bond stabilization.  Modelling realistic reaction conditions is out of the
scope of the present work.  This would typically call for the calculation of several
adsorption and coadsorption configurations, coverages and reaction paths\cite{stoltze85}.  We also
concentrate in the small molecular species B$_x$N$_y$, where x,y=\{0,1\}.  Considering bigger molecules 
at the DFT level becomes computationally very difficult as the number of the possible 
molecules increases as 2$^n$ (n=x+y).

\section{Methods}
\label{sec:methods}
\subsection{General concepts}
In \abit calculations, realistic catalyst nanoparticles are frequently modelled by
slabs in supercell geometry, consisting of 3-6 atomic layers of catalyst and a sufficient amount
of vacuum ($>$10\AA) between the slabs.  The slab usually contains  a step edge
in order to model a realistic nanoparticle with active sites \cite{logadottir03,honkala05,abildpedersen06}.
%The infinite slab consists of a periodically repeated unit cell and in the following,
%we refer to the adsorbed surface species in the unit cell with an asterisk ($^*$).

In the following, we assume that two adsorbates, X$^*$ and Y$^*$, are far away from
each other on the surface and we bring them together to form a new adsorbate species XY$^*$.  The energy for 
this reaction  X$^*$+Y$^*$$\rightarrow$XY$^*$ can be calculated as follows:
\begin{equation}\label{surfreact}
\Delta E = \big{(}E(XY^*)+E_0\big{)}-\big{(}E(X^*)+E(Y^*)\big{)}, \\
\end{equation}
where $E(X^*)$ is the energy of the adsorbed surface species X$^*$ and E$_0$ is the energy of
a surface unit cell without adsorbates.  We manipulate \eq{surfreact} as follows:
\begin{equation}\label{stage1}
% \left .
\begin{array}{c}
\Delta E = \big{(}E(XY^*)+E_0\big{)}-\big{(}E(X^*)+E(Y^*)\big{)} \\ \\
= \big{(}E(XY^*)-E_0\big{)}-
\big{(}(E(X^*)-E_0)+(E(Y^*)-E_0)\big{)} \\ \\
= E_s(XY^*)-(E_s(X^*)+E_s(Y^*))
\end{array},
% \right .
\end{equation}
in the last line of the equation, we have used energy values $E_s$ defined as:
\begin{equation}\label{surfe}
E_s(X^*) = E(X^*) - E_0.
\end{equation}
We observe from \eq{stage1}, that using ``shifted'' energy values $E_s$ defined in \eq{surfe}, 
we can calculate the reaction energy for a reaction X$^*$+Y$^*$$\rightarrow$XY$^*$ on the surface 
with the simple formula
\begin{equation}\label{surfreact2}
\Delta E = E_s(XY^*)-(E_s(X^*)+E_s(Y^*)).
\end{equation}
In the Results section, we tabulate values of E$_s$ in different parts of the catalyst
surface (terrace, edge) and then use these tabulated values to calculate reaction energetics
using \eq{surfreact2}.

Using the same notation, the adsorption energy can be written as follows:
\begin{equation}\label{ads}
E_{ads}=E(X^*)-E(X)-E_0 = E_s(X^*)-E(X),
\end{equation}
and the dissociative adsorption energy, i.e. energy for reaction XY(g)$\rightarrow$X$^*$+Y$^*$ as:
\begin{equation}\label{adsdis}
E_{dis}=E_s(X^*)+E_s(Y^*)-E(XY).
\end{equation}
where E(X) is the energy of the molecular species in the gas phase.

\subsection{Computational methods}
\label{sec:comp}
The calculations were performed with programs in the framework of
the density functional theory (DFT), as implemented in two different codes, SIESTA and VASP.
The SIESTA code \cite{soler02,sanchezportal97} uses pseudo-atomic orbitals as its basis set, while
VASP \cite{vasp1,vasp2,vasp3} is based on plane waves.  SIESTA relies on the pseudopotential method
to describe the core electrons, while projected augmented waves (PAWs)\cite{paw} can be used in VASP.
All calculations were done with periodic boundary conditions, collinear spin and using the Perdew-Burke-Ernzerhof (PBE) general gradient approximation (GGA)\cite{gga}. We use the Monkhorst-Pack (MP) sampling\cite{mp} of the Brillouin zone in calculations involving the slab.  As we are using different k-point samplings, we will indicate
the fineness of the M$\times$N brillouin zone sampling also with the area of the reciprocal
space per one sampled k-point (\abz). In this work, preliminary calculations were typically done with SIESTA, while the final energies were always calculated with VASP. Due to the more systematic control of accuracy in the VASP code, we use it as a benchmark for the more computationally efficient SIESTA code.  Nudged Elastic Band (NEB)
calculations\cite{henkelman00} for reaction barriers were performed entirely with VASP.

%During our calculations, we saw that the two different schemes gave consistent results
%in terms of energy ordering of the adsorption geometries.
% remove the energies stuff, explain CG.
%Only if the energy difference of geometries was less than [] $\sim$ eV in the approximative SIESTA
%calculation, they might have changed their relative energy position in the VASP calculation.
% ARK: Do we need the following paragraph? .. no
% We also observed that VASP reported frequently of finding a completely flat energy landscape during
% the Conjugent Gradient (CG) relaxations and the relaxation was stopped (because of this the CG relaxation had to be restarted
% several times).  On the other hand, SIESTA has considerably more noise in its forces (due to the cutoff of
% the orbitals and the grid) and it was succesfull in passing through these energy regions without restarting
% the CG optimization.

\subsubsection{SIESTA}
\label{sec:siesta}
In SIESTA calculations, Troullier-Martins \cite{tm}
scalar-relativistic pseudopotentials, with non-linear core-corrections were used. The density of the real space grid was defined
by a corresponding plane wave cutoff of $\sim$ 350 Ry and the effective density of the grid was further
increased using a grid cell sampling of 12 points. The basis set used by SIESTA consists of numerical pseudo-atomic orbitals \cite{artacho99,junquera01,soler02}.
These orbitals are obtained from the same atomic calculation that is used to generate
the pseudopotentials (thus the name ``pseudo-atomic'').  The cutoff radii and the
amount of confinement of these orbitals can be defined either by the cutoff radii (r$_c$)
or by the ``energyshift'' parameter (E$_{shift}$), larger energyshift corresponding
to increasingly confined orbitals and smaller cutoff radii \cite{sankey89}.
In SIESTA, a typical basis set is the double-$\zeta$ polarized (DZP), that consists of doubled atomic orbitals and an extra set of polarization orbitals created using perturbation theory.
A typical value for the E$_{shift}$ parameter in solids is $\sim$ 200 meV.

% 2*s + 2*3*p + 5*d = 13
For the molecular species in this study, we used the DZP basis set and E$_{shift}$=150 meV.
In the case of boron this leads to a basis set with doubled 2s and 2p-orbitals, plus an
additional set of 3d-orbitals. The total amount of orbitals is then 13 for one boron atom.
The cutoff radii defined using the energyshift for boron are 2.7 \AA ~(2s), 3.3 \AA ~(2p) and 3.3 \AA ~(3d).
For nitrogen the cutoff radii from the energyshift are 2.0 \AA~(2s), 2.5 \AA~(2p) and 2.5 \AA~(3d).

SIESTA has earlier been used to simulate iron nanoparticles \cite{izquierdo00,postnikov03} using both the
SZSP and DZSP basis sets.  The SZSP consist of 3d, 4s and 4p orbitals while in DZSP 3d and 4s orbitals are doubled.
In refs.\cite{izquierdo00,postnikov03} an explicit confinement radius of r$_c$=2.3 \AA~ for both SZSP
and DZSP basis sets was used and it was demonstrated that these basis sets with r$_c$=2.3 \AA~ produced very well
the properties of iron, including the magnetism \cite{postnikov03}.  However, in the present
case and while studying chemisorption of molecules on iron surface, we prefer longer cutoff radii
and thus use a SZSP basis with E$_{shift}$=150 meV to define the cutoff radii of the orbitals.
This way, the cutoff radii for the iron orbitals are 2.41 \AA~(3d), 3.9 \AA~(4s) and 3.9~(4p).
In our basis set, all atoms have then basis orbitals that extend at least up to 2.5 \AA~and some of them
up to 3.9 \AA.
% removed the following sentence (not relevant.. only if someone asks)
% In general, basis sets having cutoff radii of $\sim$ 3.5-4 \AA~have been demonstrated to work even with cases involving physisorption \cite{rurali05}.

% remove the parameters!
We represent the surface by a 3-layer iron slab, with the vacuum between neighboring slabs being always
$\sim$ 14 \AA.  When placing a molecule on top of this slab,
only the molecule and the top iron layer are allowed to move during the
conjugent gradient (CG) geometry optimization.
In order to speed up the calculation, the parameter adjusting the convergence
of the self-consistency cycle
%(\texttt{DM.Tolerance})
is increased to 10$^{-3}$.
This will affect the accuracy of the forces, so we simultaneously increase the
force tolerance criterion for stopping the CG relaxation
%(\texttt{MD.MaxForceTol})
to 0.1 eV/\AA.
MP sampling is chosen to be 1$\times$2, corresponding to \abz=0.15 \AA$^{-2}$.
The idea of this approximative calculation is to get a sound initial guess
for the next stage, in which we use the VASP code.

\subsubsection{VASP}
In VASP, PAWs were used.  The cutoff energy of the plane wave basis set
was always 420 eV. We represent the surface by a 4-layer iron slab, with the vacuum between neighboring slabs always $\sim$
14 \AA. Only the bottom layer is fixed to the bulk positions
during the CG relaxation.
Mixing scheme in the electronic relaxation is the Methfessel-Paxton method\cite{mpax} of order 1.
%(\texttt{ISMEAR=1}).
In a first stage, the system is relaxed using a 1$\times$2 MP sampling, which corresponds to \abz=0.16 \AA$^{-2}$.
When needed, the CG relaxation is automatically started again or until the forces have
converged to a minimum value of 0.01 eV/\AA.
After this, the relaxation is continued with MP sampling of 3$\times$5, corresponding to \abz=0.02 \AA$^{-2}$ and
CG relaxation is restarted if needed. This way we are able to reach a maximum force residual of $\approx$ 0.02 eV/\AA.
%In the final step, a single electronic minimization (no geometry optimization) is performed using the
%tetrahedron method \cite{tetra}.
%(\texttt{ISMEAR}=-5),
%which enhances the k-point sampling.
In all calculations special Davidson block iteration scheme was used
%(\texttt{IALGO}=38)
and symmetries of the adsorption geometries
were not utilized.  The standard ``normal'' accuracy was used.
%(\texttt{ISYM=0}).
%Dipole correction of the total energy to the direction of the normal of the slab was turned on.

In the case of NEB calculations, and due to the large number of atoms
we are considering, only three image points (plust the two fixed points) were used.
In general, we observed that NEB calculations with large surface slabs can be tedious;
some configurations at the lowest energy path could bring down their total energies
by shifting the iron layers in a collective movement and this way change the relative
position of the adsorbant molecule to energetically more favorable site. 
To avoid this unphysical situation, we fixed the lowermost
layers and relaxed only the topmost iron layer and the adsorbed molecule during the NEB calculations.  
This must exaggerate the reaction barriers, but we believe that this approximation should be valid for comparative estimations of the order of magnitude of the reaction barriers and for the observation of rate-limiting steps.

\begin{figure}\centering
\putgraph{width=8cm}{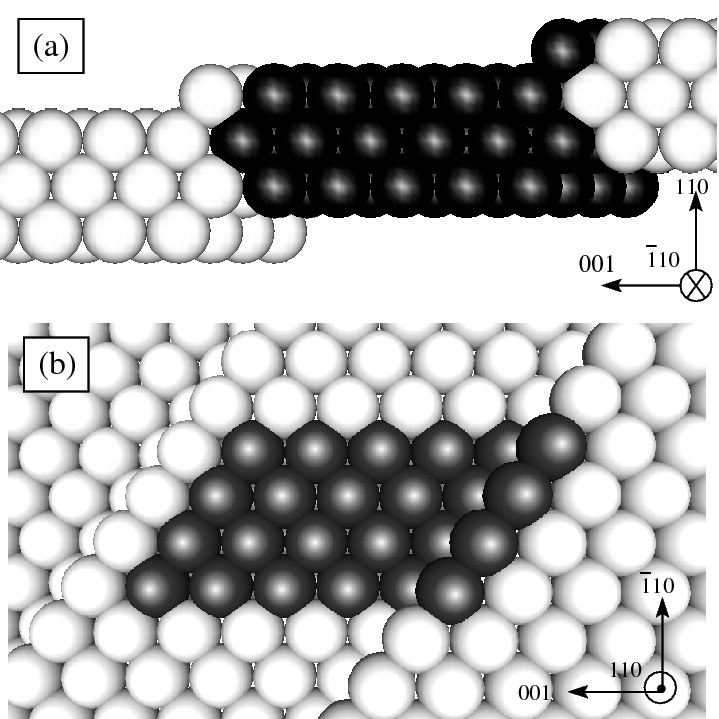}
\caption{
BCC iron (110) surface with a step.  The unit cell which was used in our
calculations is indicated by atoms with black color.  Unit cell in this
figure shows a three layer slab.  Lengths of the unit cell sides are 9.8 and 15.6~\AA.
\label{fig:unitcell}}
\end{figure}
\begin{figure*}\centering
\putgraph{width=13cm}{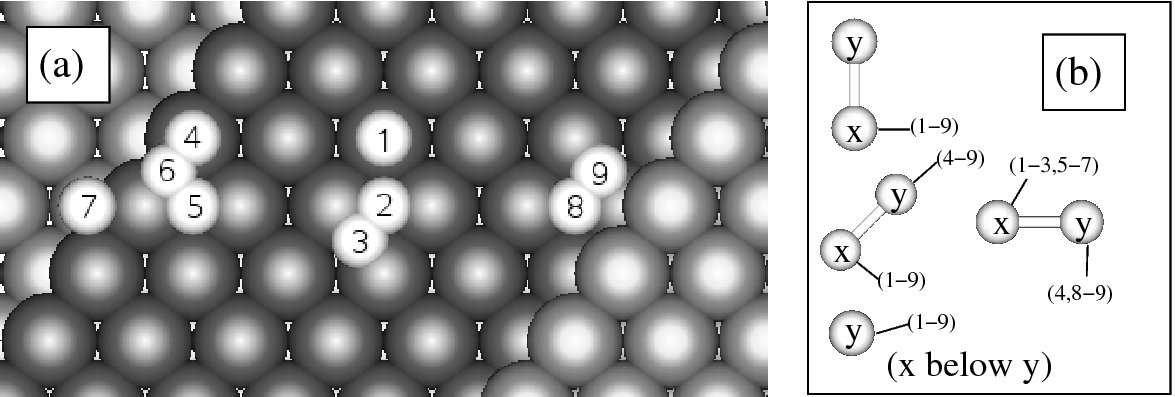}
\caption{
(a) Different sites tried out for chemisorption of molecules in the stepped
iron slab.  Sites (1-3) correspond to flat surface, while sites (4-9) are
in the vicinity of the step edge.  Sites 1 and 4 correspond to ``top'' sites,
2, 5 and 8 to ``hollow'' sites and 3, 6,7,9 to ``bridge'' sites.
(b) Different positions tried out for chemisorption of molecules.  The positions
in panel (b) have the same perspective as the surface slab in panel (a).
How these positions and sites are used to search for the optimal adsorption site, see \sct{sys}.
\label{fig:sitesnum}}
\end{figure*}

\subsection{Adsorption sites}
\label{sec:sys}
The unit cell used in our calculations is depicted in \fig{unitcell}.
The coordinates of the iron surface atoms were always scaled to the computational
lattice constant, which for SIESTA and VASP were 2.89~\AA~and 2.83~\AA,
respectively (the experimental value of the lattice constant for BCC iron
being 2.87~\AA\cite{kittel}). The unit cell of \fig{unitcell} has either 68 (3-layer slab)
or 92 (4-layer slab) atoms.  Using a large enough unit cell, including both
flat and stepped region, allows us to perform a comparative study of the
adsorption energetics near and far away from the step.  A large unit cell
should also allow for more realistic relaxation of the topmost iron atoms.
We will now explain our strategy for searching the optimal geometries
of adsorbed molecules on the surface.
% ARK: point out the step edge in the unit cell figure! ok
% ADAM: pretty obvious to me, but no harm in putting a dashed line... 

In \figa{sitesnum}{a} we are considering nine different sites.  Sites (1-3)
are in a close-packed region of the iron surface.  The remaining sites
are either on top or in the vicinity of the step edge. In \figa{sitesnum}{b} different positions of a diatomic molecule
have been considered.  For each position, a set of numbers has been associated.  This nomenclature
corresponds to the site numbering of \figa{sitesnum}{a}.  The positions together
with the associated site numbers constitute the systematic search for the adsorption site.
This procedure is more clearly understood with the example of the BN molecule:  At the beginning, we will assign the labels x and y used in \figa{sitesnum}{b} as (x=B,y=N).  After this, the BN
molecule would be positioned according to each rotation in \figa{sitesnum}{b} and
for each rotation, the atom (x=B) is placed on the sites, indicated by the numbers
for the x label in \figa{sitesnum}{b}.  As BN has two different atomic species, we must
repeat the procedure with (x=N,y=B).
For a diatomic molecule with two different species, this accounts for 66 trial configurations and for a
molecule consisting of one species only, half of that.
% ARK: explain numbers in the picture for systematic search
% ADAM: if you mean the numbers in (b), then they should be explained...in fact, I don't really get (b) at all. What is beta? x below y? maybe it is enough just to show the configurations, without all the additional detail.

We perform the systematic search described above for each atom (N, B) and for each molecule (\nn, \bb, BN),
using the approximative SIESTA calculations.  During this first stage, quite many of the different
trial configurations relax into the same energy minimum.  Some 5-10 of the most favorable
adsorption geometries are then recalculated with VASP for final results.

\section{Results}
\label{sec:results}
\subsection{Iron slab properties}
\begin{figure}\centering
\putgraph{width=8cm}{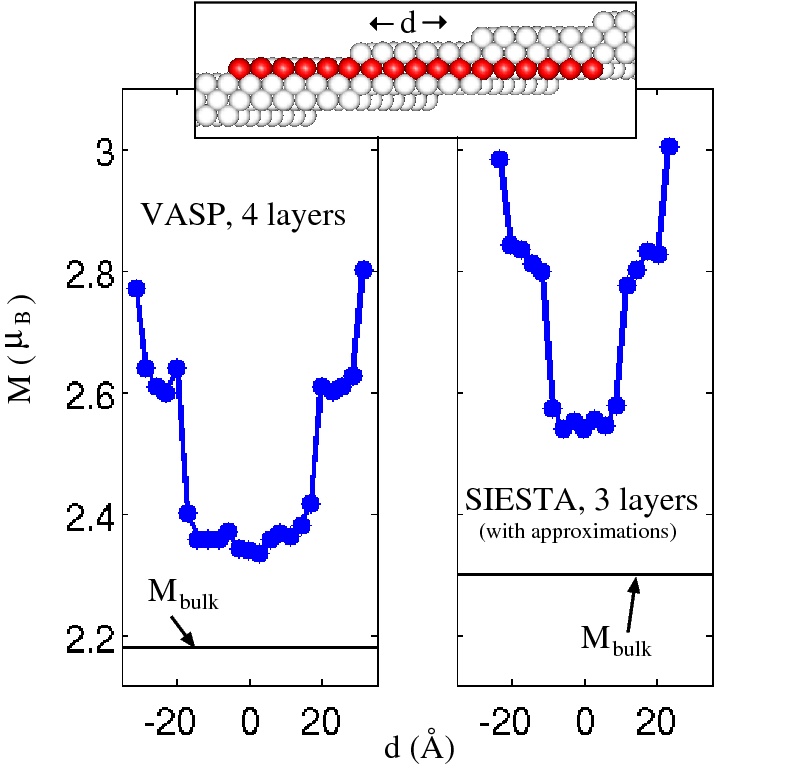}
\caption{
Magnetic profile of the stepped iron slab of \figs{unitcell}{sitesnum},
when moving along atoms indicated by red color in the topmost panel.
Left panel: magnetic profile using VASP and a 4-layer slab.  Right panel:
magnetic profile using SIESTA with 3-layer slab and some approximations 
(see \sct{siesta}). Bulk magnetism (M$_{bulk}$) has been indicated by a
solid line for both SIESTA and VASP.
\label{fig:magprof}}
\end{figure}
Magnetism is known to play an important role in iron
nanoparticles.  Typically, the magnetic moment in the
nanoparticle surface is increased, and deeper inside the nanoparticle,
the magnetic moment approaches that of bulk iron.  The central atom
of small nanoparticles might even obtain a minority spin \cite{postnikov03}.

To test for this gradual change of magnetism when approaching the
nanoparticle surface, we have plotted the magnetic profiles of the
slabs used in this work in \fig{magprof}.  In the case of SIESTA and VASP
we have used the approximations described in \sct{comp}.
For SIESTA, we obtain a bulk magnetic moment of 2.3 \mub.  Going from the center of
the slab towards surface, the magnetic moment varies from 2.5 up to 3.0 \mub.
For VASP, the bulk magnetic moment is 2.18 \mub and in the slab it varies from 2.3 to 2.8 \mub.
The experimental value for iron bulk magnetic moment is 2.2 \mub \cite{kittel}. In both cases, the atoms at the step edge obtain the highest magnetic moment.
% , a phenomenom which can be attributed to the d-band narrowing \cite{jenkins01}.
In \fig{magprof}, the magnetic profiles start
from d$\approx$-30\AA~with the high magnetic moment of the step edge atom.
The magnetic moment is lowered by $\approx$ 0.2\mub for atoms residing at the terrace.
As we move under the terrace, magnetic moment is lowered again approximatively by the same amount. SIESTA, with the SZSP basis set and the approximations described in \sct{comp},
gives slightly exaggerated magnetic moments (by $\approx$ 0.2\mub when compared
to VASP), but the overall behaviour is consistent with VASP.

In general, the magnetic moment at the top surface layer is enhanced by
20\%-30\% when compared to the bulk values.  This is consistent with
the behaviour of magnetism in iron nanoparticles\cite{postnikov03} and on
transition-metal surfaces.\cite{jenkins01}

\subsection{Reactions of molecules on the catalyst}
As we explained in \sct{setup} where we motivated our computational approach, 
we concentrate on the most simple molecules that can be formed from N$^*$ and B$^*$ that are adsorbed on the
catalyst surface and look directly at the energetic balance of the reactions \rea~that form
BN$^*$, \nn$^*$~and \bb$^*$~.  When calculating the reaction energies, we use \eq{surfreact2}
and tabulated values of E$_s$.

\begin{figure*}\centering
\putgraph{width=15cm}{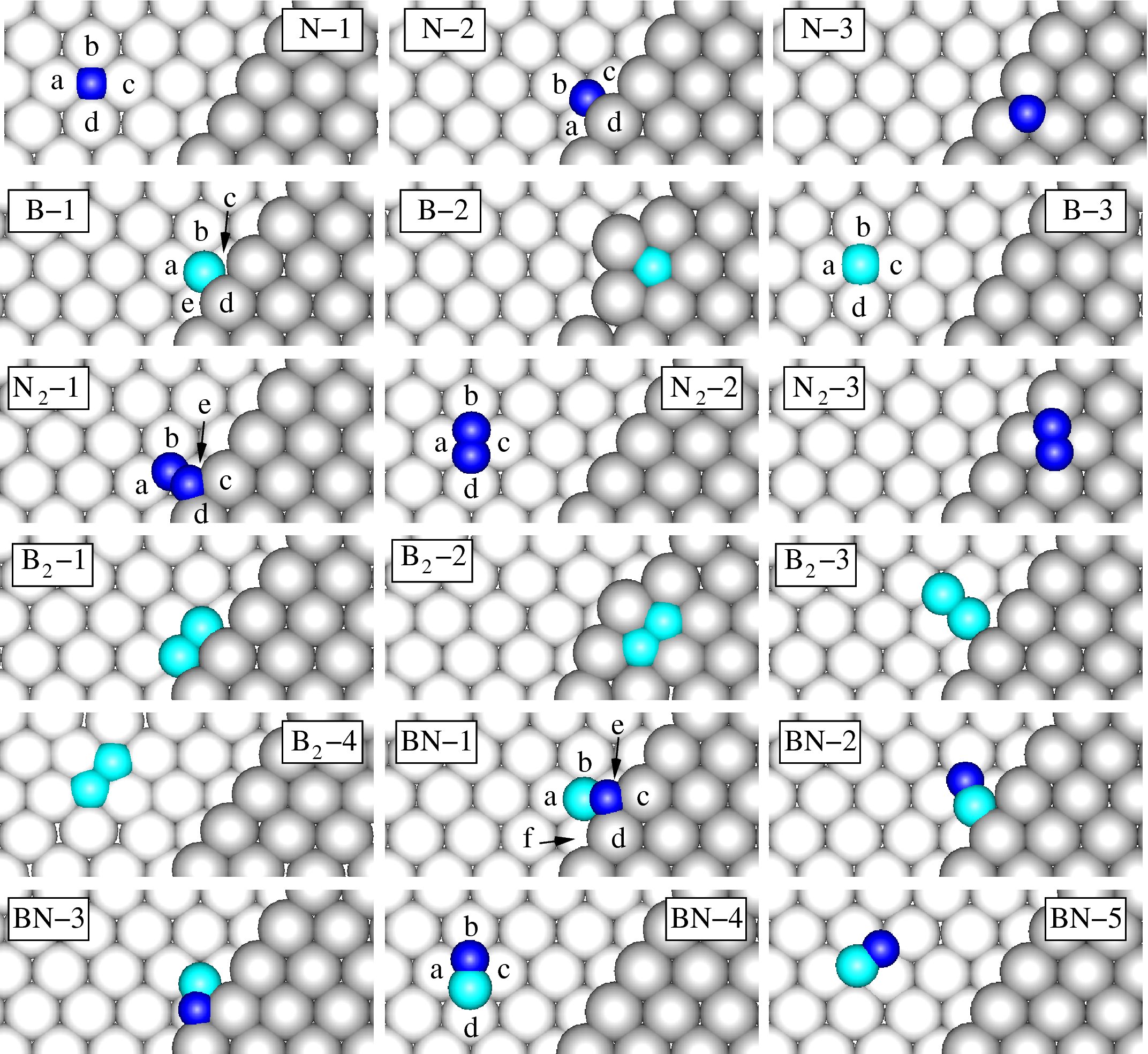}
\caption{
Some of the most stable geometries for B$_2$,BN and N$_2$ molecules and
the B and N atoms on the iron surface.  Different geometries are tagged with the
same labels as in \tab{surfnrjs}.  In the case of BN, magenta (blue) corresponds to
boron (nitrogen).
\label{fig:allgeoms}}
\end{figure*}

The optimal positions for adsorbed N, B, \nn, \bb~and BN molecules have been found using
the approach described in \sct{sys} and they are illustrated in \fig{allgeoms}.  The indices given
to these molecular geometries (\bb-1, \bb-2 etc.) are the same as used in
\tabb{surfnrjs}{surfreact} and in the density of state plots in \fig{moldos}. The main results of the adsorption energetics on the iron slab have been collected in \tab{surfreact}.  There the energetics have been categorized according to different regions of the iron slab of \fig{sitesnum}: The ``terrace'' corresponds to sites (1-3), ``edge'' region to sites (4-9) and the ``terrace and edge'' to all sites in \fig{sitesnum}.  In each class the energetically most favorable surface geometry has been considered.  In the ``terrace and edge'' column, the atoms are free to choose either
terrace or edge sites (whichever is favorable), leading to different values than in ``edge'' and ``terrace''
rows.

\begin{table}
\begin{tabular}{|l|l|l|l|}
\hline
Adsorbate & E$_{ads}$ (eV) & E$_{s}$ (eV) & BL (\AA) \\
\hline\hline
N-1&-6.6&-9.7&\\
N-2&-6.4&-9.5&\\
N-3&-6.2&-9.3&\\
N-4&-5.9&-9&\\
\hline\hline
B-1&-6.7&-7&\\
B-2&-6.6&-6.9&\\
B-3&-6.3&-6.6&\\
\hline\hline
N$_2$-1&-1.2&-17.7&1.33 (1.12) \\
N$_2$-2&-1.1&-17.6&1.28\\
N$_2$-3&-1.1&-17.6&1.29\\
\hline\hline
BN-1&-8.1&-16.9&1.4 (1.34) \\
BN-2&-7.8&-16.5&1.39\\
BN-3&-7.7&-16.4&1.43\\
BN-4&-7.3&-16.1&1.38\\
BN-5&-7.3&-16.1&1.42\\
\hline\hline
B$_2$-1&-9.9&-14.1&1.78 (1.62) \\
B$_2$-2&-9.6&-13.8&1.73\\
B$_2$-3&-9.3&-13.5&1.76\\
B$_2$-4&-9.3&-13.5&1.77\\
\hline
\end{tabular}
\caption{
Adsorption energies E$_{ads}$ and energies E$_s$ (see \eq{surfe}).  Values of E$_s$  % change! 
can be used
directly to calculate reaction energies on the surface by using \eq{surfreact2}.  Values for N$_2$, BN and B$_2$ molecules and N and B atoms in different adsorption geometries on the iron surface have been tabulated.  Bond lengths (BL) on the adsorbant and in the vacuum (in parenthesis) are listed.  Sites and geometries have the same labels as in \figs{allgeoms}{bn_b2_levels} and in \tabs{surfnrjsmove}{surfreact}. % change!
\label{tab:surfnrjs}
}
\end{table}
%N2 1.12
%NB 1.34
%B2 1.62

\begin{table}
\begin{tabular}{|l|l|l|l|l|}
\hline
Adsorbate & E$_{ads}$(t)& E$_{ads}$(e)& E$_{ads}$(t+e)&  E$_{ads}$(e) - E$_{ads}$(t)\\
\hline\hline
N\sta&-6.6&-6.4&-6.6&0.2\\
B\sta&-6.3&-6.7&-6.7&-0.5\\
N$_2$\sta&-1.1&-1.2&-1.2&-0.1\\
NB\sta&-7.3&-8.1&-8.1&-0.8\\
B$_2$\sta&-9.3&-9.9&-9.9&-0.7\\
\hline
\end{tabular}
\caption{
Adsorption energies E$_{ads}$ for N$_2$, BN and B$_2$ molecules and N and B atoms in different parts of
the iron surface.  Terrace region (t) corresponds to sites (1-3),
edge region (e) to sites (4-9) and the whole surface (t+e) to all sites in
\fig{sitesnum}.  The energy difference when moving the atom from the optimal
site at the terrace (t) to the optimal site in the edge (e) is calculated in the last column.
All energies listed are in the units of eV.
\label{tab:surfnrjsmove}}
\end{table}

\begin{table*}
\begin{tabular}{|l|l|l|l|}
\hline
Reaction & $\Delta E$ (terrace) & $\Delta E$ (edge) & $\Delta E$ (terrace and edge) \\
\hline\hline
2N\sta $\rightarrow$ N$_2$\sta&1.7 ({\scriptsize 2(N-1)$\rightarrow$\nn-2})&1.3 ({\scriptsize 2(N-2)$\rightarrow$\nn-1})&1.6 ({\scriptsize 2(N-1)$\rightarrow$\nn-1})\\

2B\sta $\rightarrow$ B$_2$\sta&-0.4 ({\scriptsize 2(B-3)$\rightarrow$\bb-4})&-0.1 ({\scriptsize 2(B-1)$\rightarrow$\bb-1})&-0.1 ({\scriptsize 2(B-1)$\rightarrow$\bb-1})\\

B\sta+N\sta $\rightarrow$ BN\sta&0.1 ({\scriptsize (B-3)+(N-1)$\rightarrow$BN-4})&-0.3 ({\scriptsize (N-2)+(B-1)$\rightarrow$BN-1})&-0.2 ({\scriptsize (N-1)+(B-1)$\rightarrow$BN-1})\\

\hline
2N\sta + 2B\sta $\rightarrow$ N$_2$\sta + B$_2$\sta&1.3&1.2&1.5\\

2N\sta + 2B\sta $\rightarrow$ 2NB\sta&0.2&-0.6&-0.4\\
\hline
\end{tabular}
\caption{
Reaction energies (eV) of some reactions on the iron surface in different
regions.  Terrace corresponds to sites (1-3), edge to sites (4-9) and the whole surface to all sites in
\fig{sitesnum}.  The adsorbate geometries that are used to calculate the energy for reaction X$^*$+Y$^*$ $\rightarrow$ XY$^*$ are indicated in parenthesis.  Geometries are tagged with the same labels (N-1, N-2, etc.) as in
\tab{surfnrjs} and \fig{allgeoms}.  Reaction energies are calculated by taking the corresponding energies 
E$_s$ from \tab{surfnrjs} and using \eq{surfreact2}.
(note: high cost for the reaction in the 4.th row is due to forcing the very unfavorable \nn~formation).
\label{tab:surfreact}}
\end{table*}

From the results of \tabs{surfnrjsmove}{surfreact}, we can conclude the following:
(1) The reaction N\sta+N\sta$\rightarrow$\nn\sta~is unfavorable in every region of the surface, (2) in the terrace,
the reaction B\sta+B\sta$\rightarrow$\bb\sta~is the most favorable, (3) in the edge region, B\sta+N\sta$\rightarrow$BN\sta~is the most favorable reaction and (4) in a situation where both terrace and edges are available, BN formation is still slightly more favorable than B$_2$ formation.
(5) All the atoms and molecules (with the exception of the nitrogen atom)
prefer to populate the step edge. 

\begin{figure*}\centering
\putgraph{width=17cm}{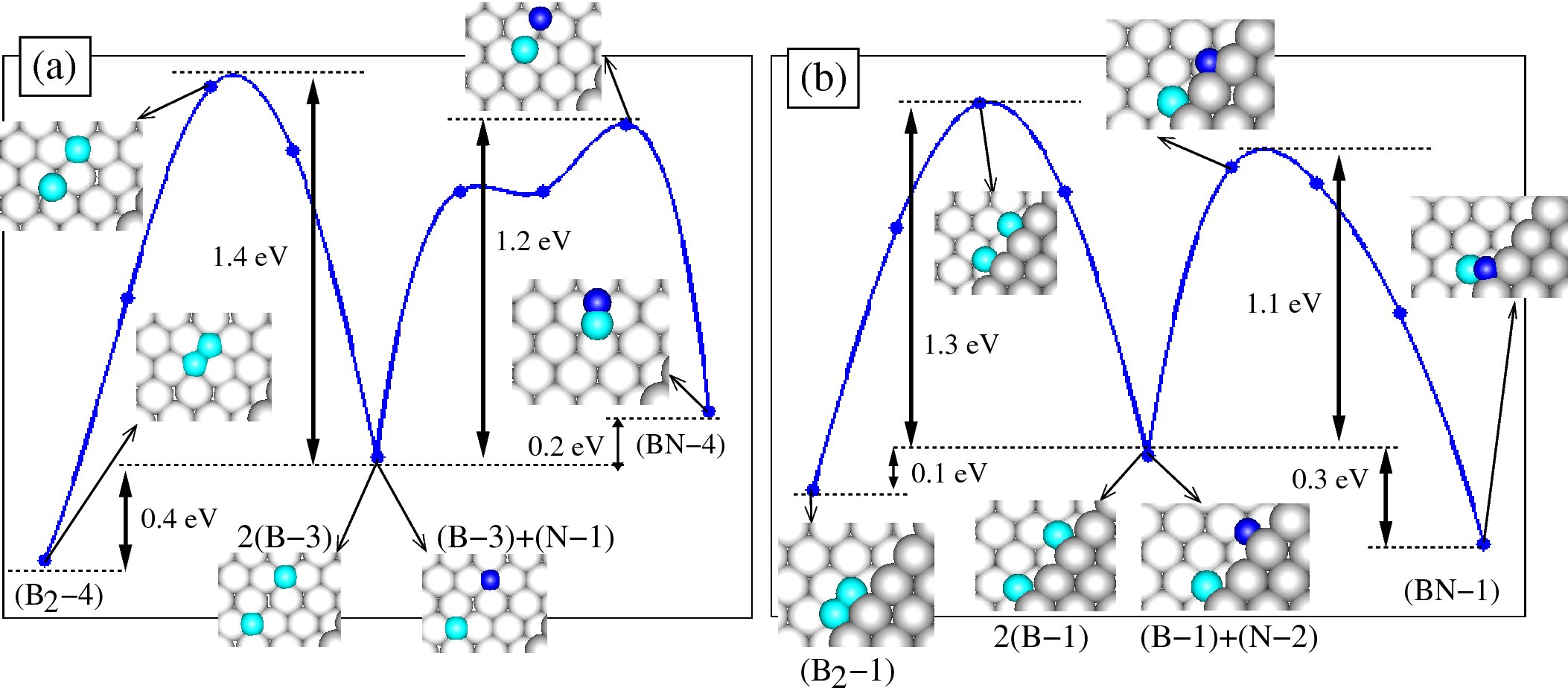}
\caption{
Reaction barriers along a few reaction paths for (a) reactions at the terrace \big{(}{\scriptsize 2(B-3)$\rightarrow$\bb-4} and  \big{(}{\scriptsize (B-3)+(N-1)$\rightarrow$BN-4}\big{)} and for (b) reactions at the step edge \big{(}{\scriptsize 2(B-1)$\rightarrow$\bb-1} and {\scriptsize (N-2)+(B-1)$\rightarrow$BN-1} \big{)}.  The slightly higher ($\approx$ 0.1 eV) energy cost for reaction  \big{(}{\scriptsize (B-3)+(N-1)$\rightarrow$BN-4}\big{)} than reported in \tab{surfreact} results from placing 
the N and B atoms in the same unit cell.
\label{fig:nebs}}
\end{figure*}

Energy barriers have been calculated along a few reaction paths for reactions 
\rea~involving boron and nitrogen both at the terrace and at the step edge.  
The reaction barriers and some atomic
configurations along the lowest energy path have been illustrated in \fig{nebs}.
From \fig{nebs} we can see that the energy barriers for competing reactions
B\sta+B\sta$\rightarrow$\bb\sta~and  B\sta+N\sta$\rightarrow$BN\sta~have the same
order of magnitude in both at the terrace and at the step edge.  No rate-limiting
steps are observed.

Next we will take a detailed look at the geometries, compare some of them
to earlier computational results and finally, based on the detailed
analysis of the geometries we give a simple explanation
why BN formations is so favorable at the step edge.  We start by looking at the adsorption
geometries of individual nitrogen and boron atoms.

% (a) terrace, (b) step edge

\subsubsection{Adsorption of N}
In the adsorption geometry N-1 of \fig{allgeoms},
changes in the positions of surface iron atoms
surrounding the adsorbed nitrogen are observed.
In order to quantify these changes, we have labelled some of the atoms
with letters a,b,c and d. The distance from the adsorbed N atom to the neighboring iron atoms a and c (b and d) is 1.79 (1.96) \AA.
Iron atoms have moved in order to create a 4-fold site for the N atom by contracting the distance b-d
by $\sim$ 5 \% and expanding distance a-c by $\sim$ 20 \%. The N atom is now almost completely incorporated
in the first iron layer and its distance from the plane formed by atoms a,b,c and d is only 0.5 \AA~
while its distance to the iron atom lying directly below is 2.47 \AA.  The rather big unit cell we
are using in our calculations has made it possible for the iron atoms to ``give way'' for the
nitrogen atom and to adsorb deeply into the adsorbant at approximately 4-fold symmetric site. In geometry N-2, the nitrogen atom has very similar coordination to N-1.  Now nitrogen has found a 4-fold site by taking advantage of the iron atoms at the step edge.  Three of the neighbour iron atoms (a,b,c) reside in the terrace, while one of them (d) sits in the step edge.  The distance of nitrogen to the nearest neighbour iron atoms are 1.87 (a), 1.90 (b), 1.86 (c) and 1.91 (d) \AA. Breaking the trend a bit, geometry N-3 prefers a 3-fold site.  This must be related to
the fact that it is in contact with two step edge atoms and so the chemical environment and
charge transfer must be different at this site.

Based on the geometries N-1, N-2 and N-3 we can conclude that, within the unit cell used in this
study, nitrogen prefers 3- or 4-fold sites with iron.  Near the step edge there is no need to adsorb deeply
into the iron layer in order to gain this desired coordination with iron.  This is particularly
true for geometry N-2 as it can easily have a 4-fold coordination with iron due to the
step edge morphology.  The energy differences between different nitrogen atom sites
are not that big.  From \tab{surfnrjs}, they are of the order of $\sim$ 0.2 eV.
From the point of view of catalytic synthesis involving nitrogen atoms, we could argue
that having more step edges than flat terrace areas on the surface is beneficial, as the adsorption
of nitrogen very deeply into the iron layer can be avoided.

In \refcite{Ref}{mortensen99_2} nitrogen adsorption on Fe(111), (100) and (110) has
been studied using DFT calculations.  It was found that on Fe(100),
nitrogen prefers a 4-fold symmetric site.  In the case of Fe(110), nitrogen was found to prefer a 3-fold
site, but the unit cell used in that case was very small and only the first-layer of iron
atoms was allowed to relax.  It was also reported that calculated adsorption energies for Fe(111) and
Fe(110) were smaller than for Fe(100), probably due to the lack of available 4-fold symmetric sites.  In our
case, an approximately 4-fold symmetric site is created in the Fe(110) surface by movement of iron atoms and the site created this way starts to resemble the one that exists in the Fe(100) surface.
It is also noted in \refcite{Ref}{mortensen99_2} that the reconstruction of iron surfaces due to nitrogen adsorption
most likely consist of geometries very similar to the one observed in Fe(100).

We also calculated a configuration where the N atom is adsorbed into a 3-fold site on the terrace
(not shown in the figures).  The adsorption of nitrogen into the 3-fold terraace site was achieved by fixing all the iron atoms in the surface slab, this way avoiding the relaxation of N into the 4-fold site (i.e. at N-1).  In this case we obtained E$_{ads}$=-6.3 eV and E$_s$=-9.4 eV.

Using a larger unit cell in our calculations would allow for stronger relaxations in the first iron layer.  
In this case, nitrogen in geometry N-1 could adsorb deeper into the adsorbant, and the situation would resemble
even more the adsorption of nitrogen into Fe(100), where the coordination of N is actually 5
(nitrogen is also bonded to the atom directly below).  However, we did not pursue this possibility, as the
computation with unit cells having $>$ 100 iron atoms is extremely heavy.
%We will discuss
%the implications of this to our results in more detail in \sct{conclusions}.

% !! ok, add here that we calculated also the 3-fold site for nitrogen with
% iron atoms fixed.. Eads=-6.3 and E_s=-9.4
% => how much is the reaction energy with this E_s ..? talk about it in the case
% of N2.

\subsubsection{Adsorption of B}
In the geometry B-1 in \fig{allgeoms}, the boron atom has quite a high coordination.  Again,
we have labelled the neighboring atoms with letters.
The distance to the nearest neighbor iron atoms are 2.03 (a), 2.48 (b), 1.93 (c), 2.1 (d) and 2.13 (e) \AA.
Distances to the iron atoms are now longer than in the case of nitrogen, but the coordination
is clearly higher.  The bigger distance comes as no surprise, due to the higher orbital radius
of boron atom when compared to nitrogen.
In general, boron is also known to prefer high coordination \cite{boustani97}. The higher coordination preference of boron is more clearly observed in the adsorption geometry B-2.
The iron step edge atoms are not as tightly bounds as the terrace atoms and for this reason the
strong reconstruction of iron atoms seen in B-2 is possible.  There are now altogether six iron atoms
surrounding the boron atom (one of them directly below the boron atom), all within a
distance of 2.0 - 2.24 \AA.

In the adsorption geometry B-3 the preference for high coordination of boron is again obvious, but
it is frustrated due to the lack of suitable sites.  No strong reconstruction, like the one seen
in geometry B-2 is observed, because arranging the iron atoms in the close-packed region would
be energetically very unfavorable.  Boron cannot push itself very deeply into the iron layer either,
the trick employed by nitrogen in N-1, as it has more extended orbital radii.  
%On the other hand, using
%a bigger unit cell for the surface slab might change the situation and allow boron to adsorb more deeply into the iron layer. 
The ``frustration'' of B-3 when compared to B-1 and B-2 is obvious in the energetics
of \tab{surfnrjs}, as B-1 and B-2 are practically degenerate and B-3 resides 0.3 eV higher
in energy.

\subsubsection{Adsorption of \nn}
Looking at the N-N bond length of geometry \nn-1 in \tab{surfnrjs}, we can see that
it has been expanded by $\sim$ 20 \%, which implies we are approaching dissociation.
In \fig{allgeoms} some of the neighboring iron atoms of the nitrogen atoms have been labelled with letters. The distances of the nitrogen atoms to their nearest iron neighbours
are 1.93 (a), 1.94 (b), 2.04 (e) \AA~and 1.9 (d), 1.95 (c), 2.12 (e) \AA.
Similar to the case of an isolated nitrogen atom, nitrogen prefers a total coordination of four
(i.e. surrounded by one nitrogen atom and three iron atoms).
It is then not surprising that \nn~prefers the step edge;  due to the morphology
of the step edge, there are sites offering 3-fold coordination with iron for each one
of the nitrogen atoms, while maintaining a reasonable N-N bond length.

The adsorption geometry \nn-2 is very similar to \nn-1 and it has N-N bond length expanded by $\sim$ 14 \%.
Now the neighboring iron atoms move, but very slightly;  the distances a-c and b-d expand
both only by $\sim$ 4 \%.  Each nitrogen atom is seen to have three iron neighbours.
The nitrogen-iron nearest neighbor distances for each nitrogen atom are 2.09 (a), 1.89 (b), 2.07 (c) \AA~
and 2.09 (a), 1.9 (d), 2.1 (c) \AA.  Again, the nitrogen atom coordination is four
(three iron atoms and one nitrogen atom). The geometry \nn-3 is very similar to \nn-1 and \nn-2 and the total energies for all adsorption geometries of \nn~molecule from \tab{surfnrjs} are almost degenerate.
The step edge geometry \nn-1 is slightly more favorable than the others, as the nitrogen
atoms can obtain their preferred coordination without significant rearragement of the iron
atoms.

Earlier calculations of \nn~adsorption on iron surface include \refcite{Refs}{mortensen99,logadottir01_2}.
In \refcite{Ref}{mortensen99}, \nn~and N adsorption on the low-coordinated Fe(111)
have been studied using DFT.  In that reference, bigger \nn~concentrations
(and smaller unit cells) were studied.  In \refcite{Ref}{logadottir01_2} the \nn~and N adsorption on Fe(110) were studied, using a 2$\times$2 unit cell, but in this study, the atoms of the iron slab
were fixed.  These earlier computational studies are therefore not directly comparable to the present
work. % (they deal with different coverage of adsorbates), but we do find several similarities.

In both \refcite{Refs}{mortensen99,logadottir01_2} the \nn~molecule was found to prefer
the ``top'' site (i.e. site (1) in \fig{sitesnum}) and a geometry where
the N-N bond projects into the vacuum (i.e. it is ``standing'' on the surface).
We also find this same adsorption geometry (not shown in \fig{allgeoms}) to be a local
minimum,
%(the nitrogen-iron distance and bond length being identical to that of \refcite{Ref}{mortensen99})
but its total energy is $\approx$ 0.6 eV higher
than that of \nn-2 in \fig{allgeoms}.  Keeping in mind that \refcite{Ref}{mortensen99} 
emphasizes that \nn~adsorption geometries where both N-atoms are in contact with the iron adsorbant
are very dependent on the coverage and that the coverage in our case is quite low, the result
we have obtained is not surprising.

\subsubsection{Adsorption of \bb}
At first sight, the adsorption geometries of \fig{allgeoms} for individual boron atoms and the \bb~molecule
are very similar. The five nearest neighbour iron atoms for a single boron atom  in \bb-1 are
within the range of 2.2 - 2.47 \AA.  The coordination of a single boron atom in \bb-1 is therefore
between 4 and 5,  which is very similar to the case of B-1.  The bond length of \bb-1 has been expanded by 10 \%. The tendency for high coordination is more clear in geometry \bb-2 where a strong reconstruction of
the iron layer, similar to the case of B-1, occurs.  For one boron atom in \bb-2 the four nearest
neighbour iron atoms are within a range of 1.94-2.32 \AA~and the total coordination of a
boron atom is then $\sim$ 5 (i.e., four iron atoms and another boron atom).

In geometry \bb-3, one boron atom resides near a step edge and has a high coordination,
while the other boron is in the terrace region and cannot get high coordination.
The boron atoms in \bb-4 have obtained high coordination through the reconstruction of the iron layer
(the situation looks very similar to \bb-2), but on the other hand, there must be a high energy cost for
moving the iron layer atoms in the close-packed region.  This can be seen in \tab{surfnrjs}, where \bb-4
lies 0.3 eV higher in energy than \bb-2.

\subsubsection{Adsorption of BN}
As we have discussed in previous sections, nitrogen and boron atoms prefer different coordination
numbers.  They maintain their preferences even when forming a molecule.
In particular, nitrogen was seen to prefer 3 to 4-fold coordination, while boron prefers
5 to 6-fold coordination.  In the case of boron nitride molecule, we should then find a suitable
surface morphology that would allow simultaneously these different coordinations for boron and nitrogen.
It is obvious that the step edge offers the best possibility for this.

Looking at \fig{allgeoms} and \tab{surfnrjs} we observe that the most favorable adsorption
sites for the boron nitride molecule are indeed at the step edge.  Looking first at BN-1, we
see that the bond length is almost equal to the free molecule, expanded only by $\sim$ 4 \%.
The nearest neighbour iron atoms for nitrogen are 1.89 (c), 1.90 (d) and 2.25 (e) \AA, while
for boron they are 1.97 (a), 2.1 (e) and 2.31 (f) and 2.43 (d) \AA.
Geometries BN-2 and BN-3 exhibit a very similar trend, i.e. the boron atom is higher
coordinated than the nitrogen atom.  The geometries BN-4 and BN-5 are almost degenerate in energy
and ``frustrated'' because the molecule is not able to obtain coordination of 3-4 for nitrogen
and 5-6 for boron due to the flat morphology of the terrace region.

\begin{figure*}\centering
\putgraph{width=17cm}{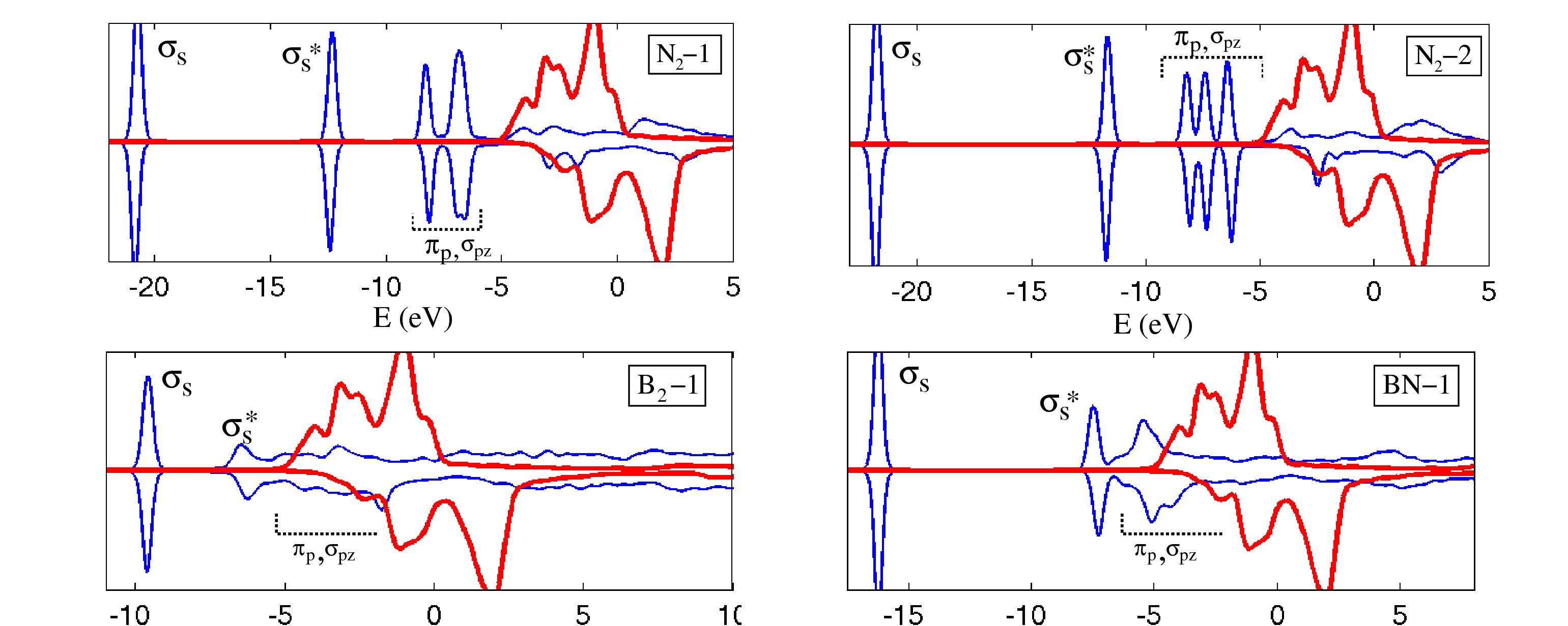} % in APS, width=17 cm OK
\caption{
Density of states, projected into atom-centered iron d-orbitals (thick red line) and into B and N atom-centered s- and p-orbitals (blue line).  The states have been interpreted using the same notation as in
\figs{n2_levels}{bn_b2_levels}.  Peaks with significant s-orbital character only when projected to atom-centered B and N orbitals are most easily identified (\sigmab, \sigmaab).  Majority (positive values) and minority spin
(negative values) are indicated.
\label{fig:moldos}}
\end{figure*}

\subsection{Electronic structure of molecules on the catalyst}
\label{sec:ele}
In this section we take a look at the electronic
structure and bonding of molecules on the iron adsorbant.  In particular, we
are interested why \bb~and BN are stabilized on the surface, while \nn~is so
unstable.  We do this by looking at the electronic states of the molecules
in vacuum and at their density of states on the adsorbant.

A classical example of this kind of analysis is the Blyholder model for the CO molecule
(see \refcite{Ref}{hu95} and references therein), where the low-lying Molecular Orbitals (MO) stay relatively inert,
while the MOs energetically near to the adsorbant d-states or overlapping with them (most
notably the HOMO and LUMO states) dominate the chemisorption energies.
Very related to our case is also the Norskov d-band model, \cite{hammer95,hammer95_2,hammer96,hammer97,hammer06}
where the metal sp-states broaden and shift the adsorbate states
and these ``renormalized'' states are then hybridized with the metal d-states.
In our case, we will take a very ``rough'' look only into the density of states without
looking at the exact details of the orbital mixing, which might be very complicated
due to the strong atomic reconstruction of the topmost iron layer
(see for example geometries N-1 and B-2 in \fig{allgeoms}).
In particular, we are interested in which type of orbitals of the adsorbate
(bonding or antibonding) interact most strongly with the metal d-states.

The iron atoms near the adsorbate are known to lower their
magnetic moments, while the adsorbate itself might be demagnetized
or even obtain a minority spin \cite{jenkins01}.  This demagnetization can also be
seen in the density of states of the adsorbates in \fig{bn_b2_levels}.

\subsubsection{Adsorption of \nn}
\begin{figure}\centering
\putgraph{width=6.5cm}{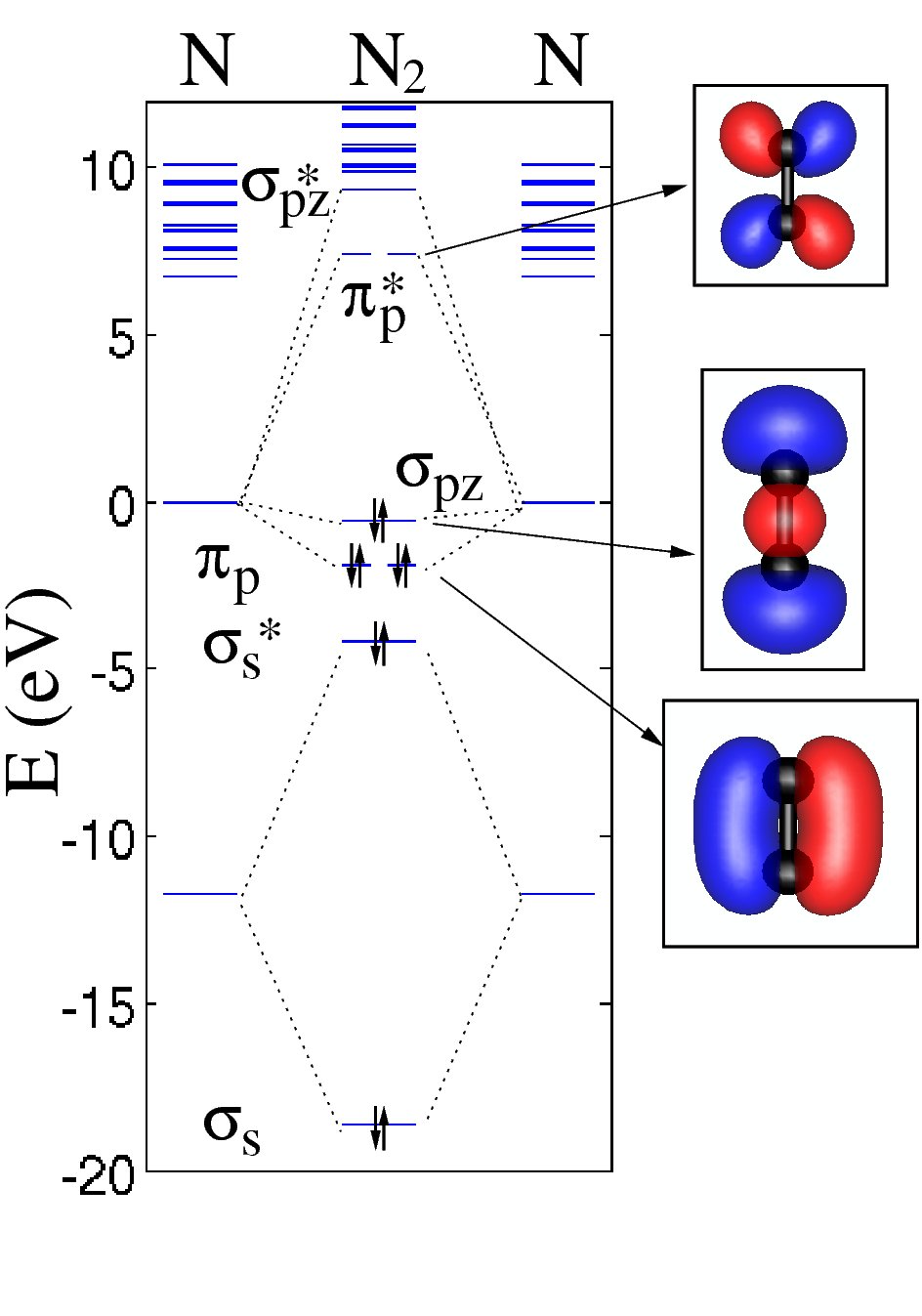} % APS: width=6.5
\caption{
(left) Energy level diagrams for individual N atoms and
the N$_2$ molecule as calculated with VASP.  N$_2$ energy levels are interpreted
using the molecular orbital theory.  The net spin-polarization of the N$_2$ molecule
is zero, so including the electron spin in the calculations does not affect the results.
Some one-electron states (from a SIESTA calculation) have been included in the insets:
color red (blue) corresponds to positive (negative) values of the wavefunction.
\label{fig:n2_levels}}
\end{figure}

The energy levels of \nn~are plotted in \fig{n2_levels} and they are
similar to earlier published ones \cite{stowasser99}.
We observe that \nn~is closed-shell and that the energy difference between
\sigmab (HOMO) and \piab (LUMO) is $\sim$ 8 eV.  The bond order of \nn~is 3, and there is no
net spin magnetic moment.  When \nn~is put in contact with an adsorbant, the bonding
is likely dominated by the $\sigma_{pz}$ and $\pi_{p}^{*}$ states.  From the
electronegativity of nitrogen and iron, we could argue that \nn~is
likely to receive electrons and thus bond through the antibonding state $\pi_{p}^{*}$ (LUMO).
To be more precise, this should depend on the relative position of the iron d-states
with respect to the renormalized \nn~energy levels, as mentioned earlier.

Comparing the PDOS graphs of \nn-1 and \nn-2 in \fig{moldos} to the energy diagram of
\fig{n2_levels}, we can easily relate different peaks to the energy levels of the isolated \nn~molecule.
In \fig{moldos} the situation is most clear in the case of \nn-2, where we find alltogether five \nn~peaks
below the iron d-states.  Two of these peaks (almost degenerate) must correspond to \pib and one to \sigmab.
There is no sign of a $\pi_{p}^{*}$ peak, so it has likely hybridized with the iron d-states.
We can then conclude that \nn~is destabilized on the iron surface through adsorption using the
antibonding \piab orbitals.

\subsubsection{Adsorption of \bb}
In \fig{bn_b2_levels} we have plotted the energy levels of a single boron
atom and the energy levels of the \bb~molecule.
We observe that \bb~has an open shell structure.  The energy difference between
\pib (HOMO) and \sigmab (LUMO) is $\sim$ 160 meV.
The bond order is 1 and \bb~has a net magnetic moment of 2 \mub.
When the calculation includes spin-polarization, an exchange splitting of the energy levels is observed
and the degeneracy of \pib orbital is removed.  Including spin-polarization in the calculation, lowers
the energy of the B$_2$ molecule by 0.84 eV.

The adsorption of \bb~is likely to happen through \pib and \sigmab orbitals, as the gap
between them is very small.
%(when compared for example to the HOMO-LUMO gap of \nn).  
Both of these orbitals are of bonding-type and this implies that \bb~will be stabilized upon
adsorption. Looking at PDOS of \bb, when it has
been placed on the iron surface (\bb-1 in \fig{moldos}), we see that both the \pib and \sigmab
MOs overlap with the iron d-states and the peaks corresponding to these MOs have
hybridized with the iron d-states.
The stabilization of \bb~on iron then looks natural
in the light of the electronic structure.
Something reminiscent of an exchange splitting
in the adsorbate PDOS peaks can be seen in the energy range from $\sim$ -4 to -1 eV.

\subsubsection{Adsorption of BN}
\begin{figure*}\centering
\putgraph{width=17cm}{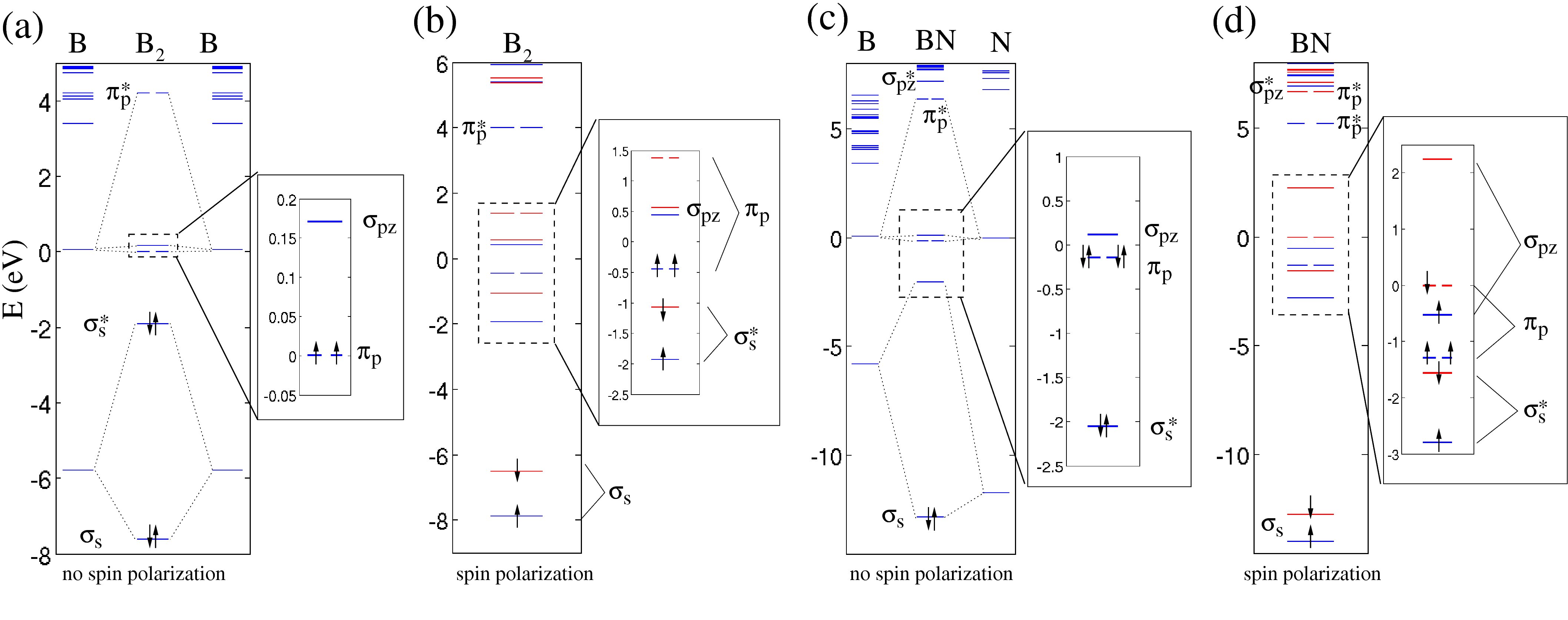} % APS: width=17 cm
\caption{
(a-b) Energy level diagrams for individual B atoms and
the B$_2$ molecule as calculated with VASP (a) without and (b) with
spin-polarization.  The B$_2$ energy levels are interpreted
using the molecular orbital theory.  (c-d) Energy level diagrams for individual B and N atoms
and the BN molecule as calculated with VASP (c) without and (d) with
spin-polarization.  The BN energy levels are interpreted
using the molecular orbital theory.  Blue (red) color corresponds to spin up (down) states.
%In the rightmost panel, some one-electron states have been marked with letters (c) and (d).
%They correspond to panels (c) and (d) in \fig{bnwfs}.
\label{fig:bn_b2_levels}}
\end{figure*}

In \fig{bn_b2_levels} we have plotted the energy levels of boron and nitrogen
atoms together with the levels of the BN molecule.
In a calculation without electron spin, the situation looks straightforward and
the BN molecule has a closed-shell structure with \pib (HOMO) and \sigmab (LUMO)
having a gap of $\sim$ 250 meV.  The bond order is 2 and there is no net spin magnetic moment. When spin-polarization is allowed, a considerable rearrangement of the MOs
due to the exchange splitting takes place: \pib and \sigmab orbitals slide through each
other in the energy-level diagram (\pib ``down'' states shift upwards, while \sigmab
``up'' states shift down) and one of the \sigmab states becomes occupied.
BN molecule lowers its energy by 0.36 eV and obtains a net magnetic moment of 2 \mub.

It is very difficult to anticipate which one of the orbitals, \pib or \sigmab,
will dominate the adsorption, as they are very close to each other in energy.  Magnetism
makes this situation even more complicated, as the gap between these molecular
orbitals can close up due to the exchange splitting.  Both of these orbitals are
of the bonding type, so at least BN should be stabilized on the adsorbant.
We look again at the PDOS plots of \fig{bn_b2_levels} and identify the peaks with the
energy levels of \fig{moldos}.  We can see that both the \pib and \sigmab states
coincide with the iron d-states and hybridize with them.
There are even some slight traces of the exchange splitting in the adsorbate PDOS peaks.
Finally, we will try to explain by means of the electronic structure only, why \bb~is more stable
on iron than BN. 

The HOMO (\pib) and LUMO (\sigmab) states for an isolated \bb~molecule in \figa{bn_b2_levels}{a} lie at energies
of 0.0 eV and $\sim$ 0.18, while for BN in \figa{bn_b2_levels}{c} they lie at $\sim$ -0.15 eV and $\sim$ 0.12 eV.
The HOMO and LUMO states of the BN molecule are then shifted slightly downwards, when compared to the same states 
of the \bb~molecule.  These states are then energetically closer to the iron d-states in \bb~than in BN.
Supporting this idea, when
looking at \fig{moldos} and comparing \bb-1 and BN-1, we can see that the
hybridization of the \pib and \sigmab states with the iron d-states seems to be more 
pronounced in the case of \bb~ and this implies that the adsorption through these bonding-type
orbitals is stronger.
%This electronic effect
%could be the underlying reason why \bb~has lower energy on close-packed iron surface
%than BN.
% If this argument is solid (it should be investigated with more care
% in the future), it will open interesting possibilities for choosing an optimum catalyst
% metal for BN synthesis by correctly situating the electronic levels of the molecules
% with respect to the density of states of the catalyst. In this case, \abit calculations
% could play an important role in choosing the optimal catalyst.
% %an optimum catalyst would have its density of states overlapping
% %by the same amount with the (renormalized) \pib and \sigmab states of \bb~and BN. 

\section{Discussion and Conclusions}
\label{sec:conclusions}
We have performed an \abit study of the energetics of the simplest chemical
reactions involved in catalytic growth of boron nitride nanotubes (BNNTs).
We studied adsorbed boron and nitrogen atoms (N\sta,B\sta)
and all their adsorbed diatomic combinations (\nn\sta,~\bb\sta~and BN\sta) on an iron catalyst.

Our objective was to study the fundamental aspect of BN bond stabilization on iron 
(rather than modelling realistic reaction conditions, see \sct{setup}).
In order to do this, we mimicked the very first stages of a CVD synthesis of BN structures.
We assumed that precursors (without defining them) have dissociated and donated individual
adsorbed N and B atoms on the catalyst.  In the very first stages of the synthesis, these atoms start 
to form either adsorbed \nn,~\bb~or BN molecules.  We believe that understanding when the BN bond is stabilized
can provide help in understanding the BNNT synthesis in general.
Specifically, we observed that \nn~is unstable, while \bb~and BN are stabilized on the iron catalyst
(BN only at the step edge region).  \nn~dissociates by adsorption on iron through antibonding 
orbitals, while \bb~and BN are stabilized by dominant adsorption through bonding-type orbitals.

On terrace regions of the iron catalyst, the reaction forming \bb~is energetically favorable, while
the reaction forming BN is not.  The energy barriers of the two competing reactions \bbrea~and \bnrea~are 
the same order of magnitude for the two reactions.  This implies that if B and N atoms are distributed on a flat iron surface, in the very first stages of the synthesis, large amounts of \bb~and individual nitrogen atoms adsorbed on iron will be formed, while very little BN molecules will form.
If further boron cluster formation occurs, it is probably not favorable from the point of view of BNNT synthesis (as mentioned in \sct{bntsyn}, BNNT growth from boron has been observed only in very elevated
temperatures).

The situation looks much more promising in the step edge region; 
the energetic balance is tipped into favor of BN formation and the energy barriers are again the same 
magnitude for both reactions \bbrea~and \bnrea.  This implies that when B and N atoms are distributed into the step edge, some \bb~molecules and considerable amount of BN molecule formation takes place.  The formation of a large number of BN molecules on the catalyst could be very important for BNNT formation, provided that these molecules are mobile and do not poison the catalyst.

%In this work we were mainly concerned about the reaction energies and if the BN bond
%can be stabilized at all on an iron catalyst.  We wanted to study why the synthesis using
%transition metals has proven to be so difficult and if there are situations where the BN
%bond is stabilized.  The emphasis is on the fundamental physics of BN bond stabilization.
%As we commented in \sct{setup}, we are far from realistic reaction conditions

%There are still many issues that should be addressed in order to make the
%picture of BN formation on iron more complete; diffusion of B and N atoms and BN molecules in both the terrace 
%and the step edge should be studied; 
%
%According to our results, BN formation on the step edge is in principle
%possible (provided there are no other serious problems not explored in this work).

The stabilization of BN at the step edge can be explained in terms of atomic coordination: 
we observed that, within the computational unit cell we used, nitrogen preferred 3-4, while 
boron 5-6 fold coordination with iron and the only morphology where these two coordinations are simultaneously available, is found at the step edge.

Summarizing, according to our calculations, the BN bond is stabilized in step edge regions of the
iron catalyst.  This implies that the yield of BNNT in a CVD synthesis might be enhanced by altering the iron 
catalyst morphology to include more steps, instead of close-packed surface regions.
Simply having step edges is not enough; having step edges, but long terraces, will result in more flat surface
sites than step edge sites, lowering the free energy for flat surface sites.
From the point of view of maximizing the BNNT yield, the terraces should then be very short.
As creating a catalyst nanoparticle with a desired morphology is very difficult,
the predictions on BN yield given in this theoretical work could be put to test in practice by using 
as a catalyst a high-index Fe surface with very short steps.

\section{Acknowledgements}
We wish to thank the Center for Scientific Computing Helsinki, for use of its computational resources.
This work has been supported in part by the European Commission under 
the 6 Framework Programme (STREP project BNC Tubes, contract number NMP4-CT-2006-03350)
and the Academy of Finland through its Centre of Excellence programme (2006-2011).

%\bibliographystyle{apsrev-oma} % aps custom bibliography style (omitting article titles and dois, etc.)
% \bibliographystyle{elsarticle-oma} % elsevier custom bibliography style (omitting article titles and dois, etc.)
%\bibliography{paper}

\end{document}